# High-Voltage Field Effect Transistors with Wide-Bandgap $\beta$-Ga$_2$O$_3$ Nanomembranes

Wan Sik Hwang[1,5], Amit Verma[1], Hartwin Peelaers[2], Vladimir Protasenko[1], Sergei Rouvimov[1], Huili (Grace) Xing[1], Alan Seabaugh[1], Wilfried Haensch[3], Chris Van de Walle[2], Zbigniew Galazka[4], Martin Albrecht[4], Roberto Fornari[4], and Debdeep Jena[1]

[1]Department of Electrical Engineering, University of Notre Dame, Notre Dame, IN 46556, USA
[2]Materials Department, University of California Santa Barbara, CA 93106, USA
[3]IBM T. J. Watson Research Center, Yorktown Heights, New York 10598, USA
[4]Leibniz Institute for Crystal Growth, Max-Born Str., D-12489, Berlin, Germany
[5]Department of Materials Engineering, Korea Aerospace University, Gyeonggi, 412791, Korea

e-mail: djena@nd.edu & whwang@kau.ac.kr

**Abstract**

Nanoscale semiconductor materials have been extensively investigated as the channel materials of transistors for energy-efficient low-power logic switches to enable scaling to smaller dimensions. On the opposite end of transistor applications is power electronics for which transistors capable of switching very high voltages are necessary. Miniaturization of energy-efficient power switches can enable the integration with various electronic systems and lead to substantial boosts in energy efficiency. Nanotechnology is yet to have an impact in this arena. In this work, it is demonstrated that nanomembranes of the wide-bandgap semiconductor gallium oxide can be used as channels of transistors capable of switching high voltages, and at the same time can be integrated on any platform. The findings mark a step towards using lessons learnt in nanomaterials and nanotechnology to address a challenge that yet remains untouched by the field.

High-voltage switching transistors are essential for a multitude of electronic applications that require the transport, conversion, conditioning, and distribution of electrical power [1]. These applications range from microscale on-chip power management for microprocessors and microcontrollers, to heavy-duty electric motor drives in automotive applications. Power transistor switches are typically large devices [1]. Large length dimensions are necessary for sustaining high voltages ranging from 100s to 1000s of volts. For a constant electric field, sustaining a larger voltage requires a thicker layer. This is because the critical *electric fields* that can be sustained in semiconductors are limited by their bandgaps through electrical breakdown processes. Traditionally silicon power transistors have been the most widely used [2]. But the $E_g$~1.1 eV bandgap and its associated critical breakdown field restricts the miniaturization, efficiency, and voltage handling capability of Si power transistors. These limitations have led to increased interest and adoption of power switching devices using wider bandgap semiconductors such as SiC ($E_g$~3.3 eV) and GaN ($E_g$~3.4 eV) [3]. The wider bandgaps and corresponding higher critical breakdown fields allow higher voltage handling capabilities, and a number of efficiency improvements result from scaling to smaller dimensions. Though wide-bandgap semiconductor devices are finding increased adoption, bulk single crystal substrates remain very expensive for SiC, or challenging for GaN.

A recent breakthrough in the fabrication of bulk single-crystal substrates of a new wide-bandgap semiconductor $\beta$-Ga$_2$O$_3$ has attracted much attention in this regard [4]. $\beta$-Ga$_2$O$_3$ has a bandgap of ~4.9 eV, significantly larger than GaN and SiC. Combined with a relatively inexpensive bulk growth method, it pushes the definition of wide bandgap semiconductors to a new level. Taking advantage of controlled doping in epitaxial layers grown on the bulk substrate, the first transistor was reported recently [5].  Because of its wide bandgap, Bulk $\beta$-Ga$_2$O$_3$ substrates can also find applications in deep-ultraviolet photodetectors [6] and light-emitting diodes (LEDs) [7].  Though the bandgap of $\beta$-Ga$_2$O$_3$ is large, it has a low thermal conductivity of 13 Wm$^{-1}$K$^{-1}$ [8] compared to Si (~150 Wm$^{-1}$K$^{-1}$) [9], GaN (~150-200 Wm$^{-1}$K$^{-1}$) [10], and SiC (360-400 Wm$^{-1}$K$^{-1}$) [11].  For transistors switching high voltages and currents, a high thermal conductivity semiconductor is desirable for efficient dissipation of excess heat in the switching process.  It is imperative to explore ways to counter the low thermal conductivity of $\beta$-Ga$_2$O$_3$ for such applications.

Further, for on-chip power management applications, it would be desirable to investigate the possible integration of wide-bandgap semiconductors on existing platforms.  Recent methods of integration of layered semiconductors such as graphene by layer transfer [12], and of GaN by direct growth on Silicon [13] points towards similar opportunities with other materials.  In this work, we find that rather surprisingly, $\beta$-Ga$_2$O$_3$ membranes of nanoscale thickness could be easily exfoliated from their bulk crystals, similar to the exfoliation of graphene and other layered crystal materials.  We report the first demonstration of high-voltage $\beta$-Ga$_2$O$_3$ nanomembrane transistors on a Silicon platform.  We compare these transistors with similar transistors fabricated from layered MoS$_2$ semiconductor channels to highlight their high-voltage handling capability.  This initial demonstration is intended to motivate research towards the integration of high-voltage transistors with various platforms using nanoscale slivers of wide-bandgap semiconductors.  Though nanoscale materials such as nanotubes [14], graphene [12], and transition metal dichalcogenides [15-18] have been actively studied for *low-power* electronic switching applications, the initial findings reported here show that nanoscale materials can also play a role in *high-voltage* electronics.

*β-Ga$_2$O$_3$* single crystals were obtained by the Czochralski method with use of an iridium crucible. The melting point of *β*-Ga$_2$O$_3$ is around 1820°C. A dynamic, self-adjusting growth atmosphere was used to minimize the decomposition of *β*-Ga$_2$O$_3$ and oxidation of iridium at the same time, as described in detail in ref. [4]. The crystal structure of *β*-Ga$_2$O$_3$ belongs to the monoclinic system with the space group C2/m and lattice constants a=12.214 Å, b=3.0371 Å, c=5.7981 Å, and *β*=103.83° [19]. It contains two crystallographically inequivalent Ga positions, one with tetrahedral geometry (Ga1) and one with octahedral geometry (Ga2). The octahedra share edges to form double chains parallel to the b-axis, which are connected by corner-sharing tetrahedral [19]. Such structure forms two cleavage planes parallel to (100) and (001) planes. Single crystals of 20 mm diameter were grown along the b-axis, i.e. parallel to both (100) and (001) cleavage planes. From the single crystal, an oriented cube of 1x1x1 cm$^3$ size was prepared with an exposed (100) plane for easy-cleaving or exfoliation purpose. Hall effect measurements in the van-der-Pauw configuration with use of In-Ga ohmic contacts showed a free electron concentration = 5.5 x10$^{17}$ / cm$^{-3}$, resistivity = 0.1 Ωcm and electron mobility = 112 cm$^2$ V$^{-1}$ s$^{-1}$, all at room temperature.

Nanomembranes of ~20 to 100 nm thickness could be mechanically exfoliated from the bulk crystal, much like the exfoliation of graphene and other transition-metal dichalcogenide layered crystals [12]. This is surprising, because *β*-Ga$_2$O$_3$ is a 3D crystal. Mechanical exfoliation is possible in layered materials due to the highly *anisotropic* chemical bonding characterized by strong in-plane covalent bonds and much weaker interlayer van der Waals bonds. Though *β*-Ga$_2$O$_3$ is not a layered material in the conventional sense, it has a monoclinic structure, with a much larger lattice constant in the (100) direction. The ease of exfoliation could be attributed to this property, though further careful study is necessary to confirm this hypothesis.

The Energy-dispersive X-ray (EDX) spectrum of a nanomembrane transferred onto Cu TEM grids is shown in Fig. 1(a). It reveals the presence of Ga and O and no other impurities. The absorption spectrum of the $β$-$Ga_2O_3$ (100) nanomembrane is shown in Fig. 1(b). The extracted optical band gap matches the expected value from the band structure calculated by hybrid density functional theory [20, 21] as shown in Fig. 1(c). Figure 1(c) shows that the conduction-band minimum is at Γ, with a small and almost isotropic electron effective mass of 0.28 $m_e$. The band gap is indirect, with a value of 4.85 eV and the valence-band maximum located along the I-L line. The direct band gap at Γ is only slightly larger, at 4.88 eV; the absorption spectra shows both the direct and indirect signatures. Calculations of matrix elements indicate that direct optical transitions at Γ are allowed, consistent with the experimentally observed sharp absorption onset [22]. In addition to the small effective mass, another feature that renders $Ga_2O_3$ attractive as an electronic material is the fact that secondary conduction-band minima are at least 2.6 eV higher in energy. The EDX spectrum and the absorption spectrum of the (100) $β$-$Ga_2O_3$ nanomembrane thus reveals that the bulk properties are preserved in the nanomembrane.

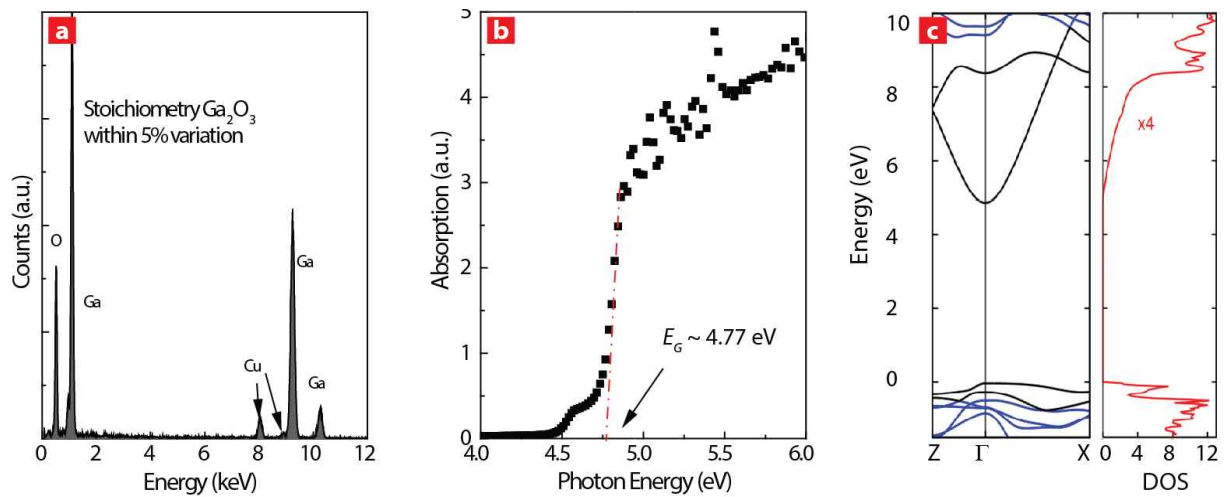

FIG. 1. (a) Normalized intensity of the Energy-Dispersive X-ray spectroscopy (EDX) spectrum for β-Ga$_2$O$_3$ nanomembrane. It shows that the β-Ga$_2$O$_3$ is formed of stoichiometry structure within 5% variation and the Cu signal is from grid of TEM. (b) Optical absorption spectra of β-Ga$_2$O$_3$ nanomembrane *vs.* photon energy, indicating bandgap of ~4.77 eV. The absorption is represented by $(\alpha h v)^2$. (c) Calculated band structure of Ga$_2$O$_3$, plotted along high-symmetry lines in the Brillouin zone. The coordinates of the Z and X points are (0, 0, ½) and (0.266, -0.266, 0), expressed as fractions of the reciprocal lattice vectors. The highest two valence bands and lowest two conduction bands are highlighted. The right panel shows the density of states (DOS); values for the conduction bands are magnified by a factor of 4 for better visibility.

In order to further investigate the structural properties of the β-Ga$_2$O$_3$ nanomembrane, transmission electron microscopy (TEM) was performed. Figure 2(a) shows the HRTEM image of the bc-plane view, the major cleavage plane of β-Ga$_2$O$_3$. The electron diffraction pattern of the bc-plane is shown in Fig. 1(b), revealing the lattice symmetry and the lattice parameters of the bc-plane. The lattice parameters of the [100] a-direction can be observed from Fig. 2(c), which is the cross-section HRTEM image of Fig. 4.

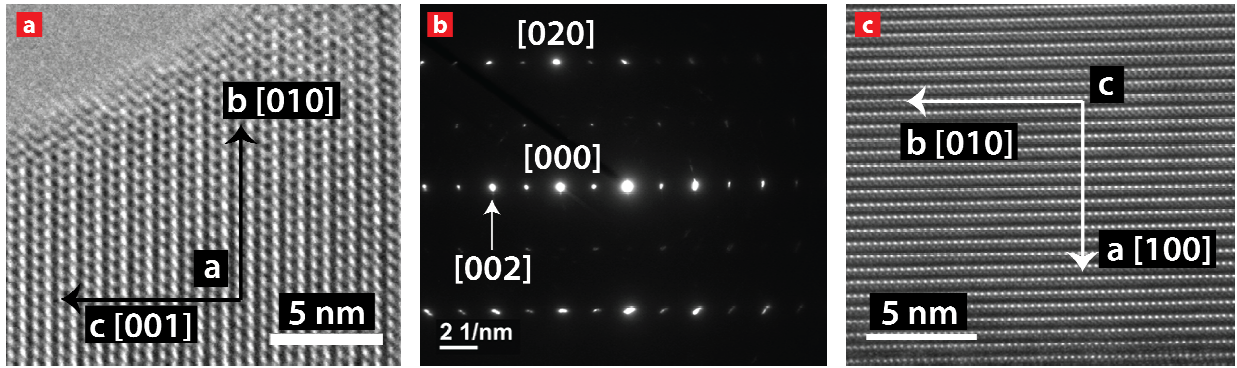

FIG. 2. TEM images of β-Ga$_2$O$_3$ nanomembrane (a) HRTEM image of bc-plan-view. (b) Corresponding electron diffraction pattern of (a). (c) Cross-sectional HRTEM view of β-Ga$_2$O$_3$ nanomembrane FET.

TABLE I. Comparison of bulk lattice parameter [19] and nanomembrane lattice parameter. The value of the nanomembrane lattice parameter is obtained from FIG 2.

|  | Lattice parameter (nm) | | |
| --- | --- | --- | --- |
|  | a (100) | b (010) | c (001) |
| Bulk | 1.22 | 0.30 | 0.58 |
| Nanomembrane | 1.22 | 0.30 | 0.57 |

The atomic properties of $\beta$-$Ga_2O_3$ nanomembrane were compared with those of bulk in Table I. It shows that the atomic properties of $\beta$-$Ga_2O_3$ nanomembrane are identical to those of bulk, indicating that no deformation or stress is introduced in the nanomembrane during the processing.

Based on the information above, $\beta$-$Ga_2O_3$ nanomembrane high-voltage FETs were fabricated as shown in Fig. 3. Nanomembranes were mechanically exfoliated and transferred onto a back-gated Silicon substrate. The source and drain contacts were defined by electron beam lithography (EBL) using Ti/Au (5/150 nm) metal stacks. The final device went through an annealing process in Ar/$H_2$ at 300 $^o$C for 3 hours to reduce the contact resistance.

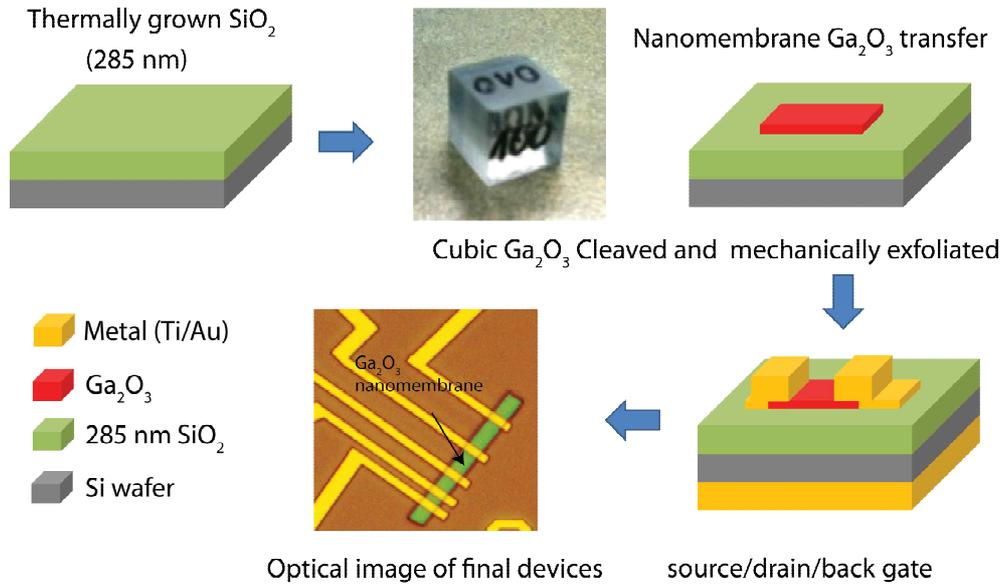

FIG. 3. Schematic process flow for nanomembrane $\beta$-Ga$_2$O$_3$ field-effect transistors. The nanomembrane thicknesses are in the range of 20 to 100 nm

Figure 4 shows the cross-section sketch of the back-gated transistor, and the corresponding TEM images of various regions of the device. The highly crystalline nature of the $\beta$-Ga$_2$O$_3$ channel region, and atomically smooth interfaces at the bottom with SiO$_2$ and on the top with the Ti/Au ohmic metal is evident in this picture. The lattice parameters of the $\beta$-Ga$_2$O$_3$ nanomembrane is identical in the channel, under the metal contacts and near SiO$_2$, and all of them are identical to the bulk value, indicating minimal strain and damage in the transfer and device fabrication process.

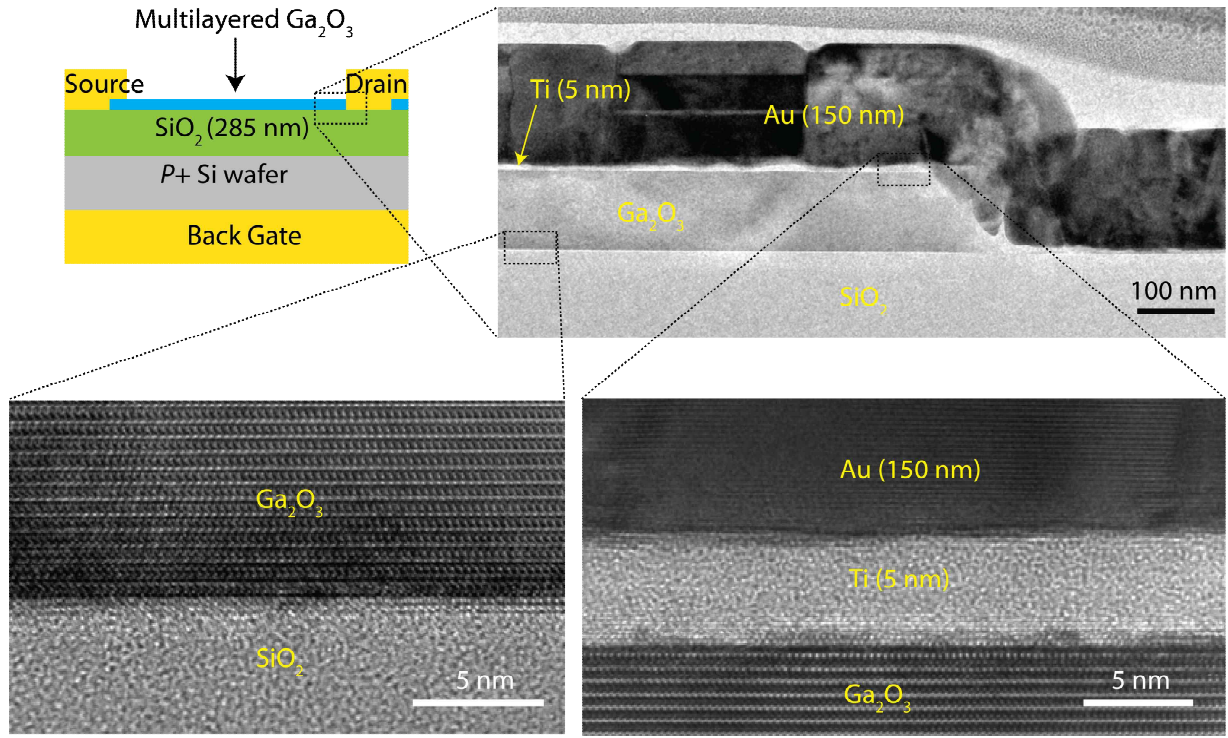

FIG. 4. Cross-sectional TEM image of $\beta$-Ga$_2$O$_3$ FETs, showing a flat interface between $\beta$-Ga$_2$O$_3$ and the SiO$_2$ dielectrics as well as between the $\beta$-Ga$_2$O$_3$ and the Ti/Au electrode.

Figure 5(a) shows the drain current versus gate-source bias, $I_D$ vs. $V_{GS}$, at room temperature at a high drain bias $V_{DS}$ = 20 V. Nanomembrane $\beta$-Ga$_2$O$_3$ FETs show a high gate modulation of ~10$^7$x even under a high drain voltage of 20 V; this modulation is limited not by the channel material, but the gate leakage current. The high current modulation is attractive for high-power and high-voltage device applications. From the transfer characteristics, we extract an "extrinsic" field-effect mobility of ~ 70 cm$^2$/Vs as shown in Fig. 5(b). This value represents a lower limit because it is not corrected for the contact resistance. The real electron mobility is expected to be around ~300 cm$^2$/V.s – closer to the bulk value. A low contact resistance can thus boost the current drive significantly [5]. The subthreshold swing (SS) of the device shown in Fig. 5(c) approaches ~200 mV/dec. The low SS, though far from the ideal 60 mV/decade is encouraging, considering the unoptimized interfaces and the thick 285 nm $SiO_2$ back-gate dielectric.

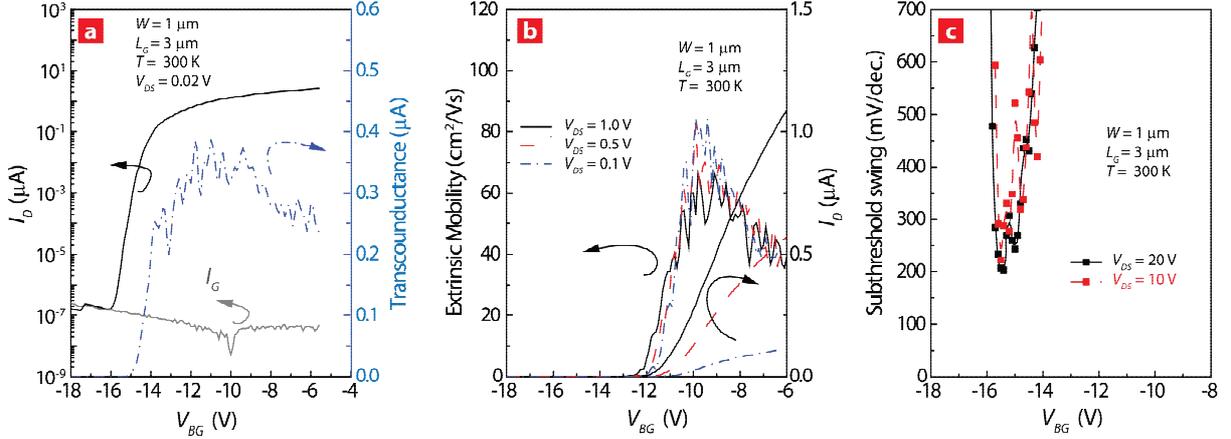

FIG. 5. Transport properties of β-$Ga_2O_3$ field-effect transistor with W/L = 1/3 μm. (a) Drain current, $I_D$ vs. back gate-to-source voltage, $V_{BG}$, showing ~$10^7$ on/off current ratio and n-type semiconductor behavior. (b) Field-effect mobility and (c) subthreshold swing vs. $V_{BG}$.

The family of $I_D$-$V_{DS}$ curves at various $V_{BG}$ in Fig. 6(a) shows typical transistor performance characterized by linear behavior at low $V_{DS}$ and current saturation at high $V_{DS}$. The contact resistance was extracted to be ~ 55 Ωmm at low $V_{DS}$. The value is comparable to $MoS_2$ FETs with an identical process flow [23], but is much higher than that of conventional Si and III-V FETs of ~0.1 Ωmm [24]. The contact resistance can be further lowered by using low work function metals and by increasing the electron density by local ion-implantation doping under the contacts. In Fig 6(a), the transistor characteristics of the β-$Ga_2O_3$ FET is compared to a multilayer $MoS_2$ channel FET of identical geometry and dimensions. Owing to the much larger bandgap, the 1 μm gate length β-$Ga_2O_3$ channel FET shows a much higher voltage handling capability than the $MoS_2$ channel FETs. In this comparison, both $MoS_2$ and β-$Ga_2O_3$ channel FETs went through the same process flow with identical gate stack structure. Fig 6(b) shows the bandgaps, and band alignmments of the two channel materials. The output characteristics of the nanomembrane $\beta$-Ga$_2$O$_3$ maintain a robust current saturation up to 70 V with no signs of output conductance. In comparison, the sharp increase in the output current of the MoS$_2$ channel FET around 10-15 V drain bias is a characteristic signature of avalanche or impact-ionization breakdown, as is expected of semiconductors of comparable bandgaps. This result shows that nanomembranes $\beta$-Ga$_2$O$_3$ channel transistors can sustain and switch high voltages even when integrated in thin layer forms on foreign substrates. High thermal conductivity but electrically insulating layers such as AlN or BN can be used to help circumvent the low thermal conductivity of the $\beta$-Ga$_2$O$_3$ channel. The high thermal conductivity insulating layers can also serve as the gate insulator for the transistor.

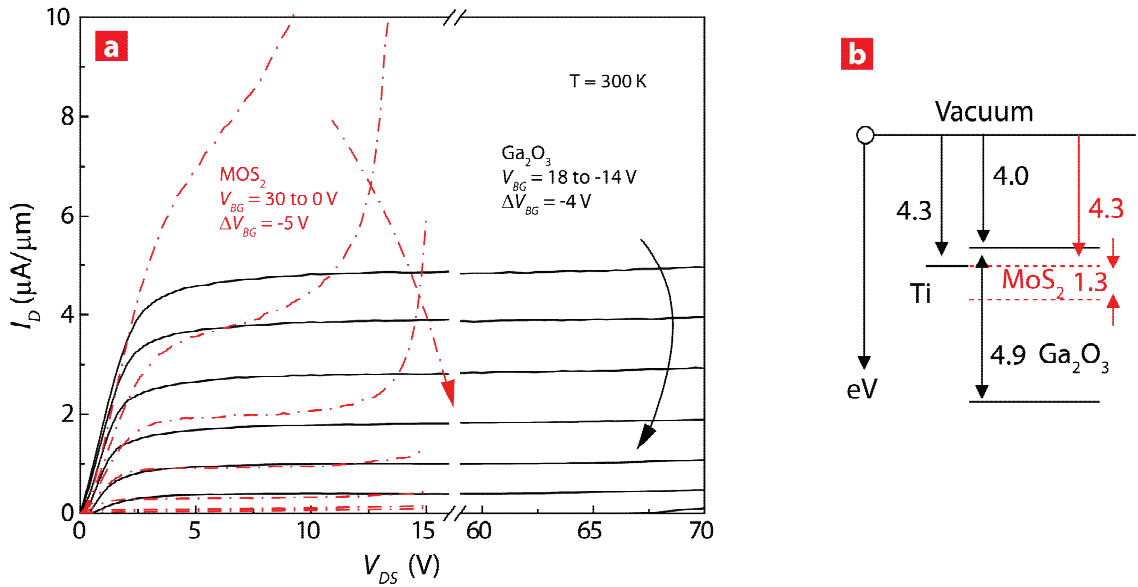

FIG. 6. (a) Common-source transistor characteristics, $I_D$ vs. $V_{DS}$ linear region and current saturation under high $V_{DS}$, and comparison of breakdown voltage of Ga$_2$O$_3$ with W/L = 1/3 μm and MoS$_2$ with W/L = 1/3 μm. (b) Band diagram of $\beta$-Ga$_2$O$_3$ comparing with MoS$_2$, indicating the formation of Schottky barrier contact between metal and $\beta$-Ga$_2$O$_3$.

In summary, nanomembrane high-voltage FETs with $\beta$-Ga$_2$O$_3$ channels were fabricated and characterized for the first time. The large bandgap of ~4.9 eV leads to high on/off ratio and high voltage transistors with low subthreshold slopes. The high breakdown field of $\beta$-Ga$_2$O$_3$ allows significantly higher voltages to be switched by orders of magnitude compared to lower-bandgap layered MoS$_2$ transistors. Since $\beta$-Ga$_2$O$_3$ has been found to have a rather poor thermal conductivity, the nanomembrane topology offers opportunities for efficient heat removal. A primary challenge is the formation of ohmic contacts. Mechanical exfoliation is indeed not a scalable method, but methods similar to smart-cut technology [25] used in Silicon-on-insulator (SOI) wafer manufacture can potentially enable controlled release of large nanomembranes of the wide bandgap material. Such a method can potentially enable the integration of nanomembrane high-voltage transistors on multiple platforms for high-voltage switching and power management.

This work was supported by the Semiconductor Research Corporation (SRC), Nanoelectronics Research Initiative (NRI) and the National Institute of Standards and Technology (NIST) through the Midwest Institute for Nanoelectronics Discovery (MIND), STARnet, an SRC program sponsored by MARCO and DARPA, and by the Office of Naval Research (ONR) and the National Science Foundation (NSF).